\documentclass[runningheads]{llncs}

\usepackage{graphicx}
\usepackage{amsmath}
\usepackage{xcolor}
\usepackage{amsfonts}
\usepackage{hyperref}

\newcommand{\modif}[1]{}
\renewcommand{\modif}[1]{{\color{black}{#1}}}
\begin{document}
\title{Cascaded multitask U-Net using topological loss for vessel segmentation and centerline extraction\thanks{This work was supported by the French \emph{Agence Nationale de la Recherche (Grant ANR-20-CE45-0011)} as part of the PreSPIN project. This work was granted access to the HPC resources of IDRIS under the allocation 2022-AD011013610 made by GENCI}}
%
%
\author{Pierre Rougé\inst{1,2} \and
Nicolas Passat\inst{1} \and
Odyssée Merveille\inst{2}}
\authorrunning{P. Rougé et al.}
%
\institute{Université de Reims Champagne Ardenne, CReSTIC, Reims, France \and
Univ Lyon, INSA‐Lyon, Université Claude Bernard Lyon 1, UJM-Saint Etienne, CNRS, Inserm, CREATIS UMR 5220, U1294, F‐69100, Lyon, France
\email{pierre.rouge@creatis.insa-lyon.fr}}
\maketitle              
\begin{abstract}
Vessel segmentation and centerline extraction are two crucial preliminary tasks for many computer-aided diagnosis tools dealing with vascular diseases.
Recently, deep-learning based methods have been widely applied to these tasks.
However, classic deep-learning approaches struggle to capture the complex geometry and specific topology of vascular networks, which is of the utmost importance in many applications.
To overcome these limitations, the clDice loss, a topological loss that focuses on the vessel centerlines, has been recently proposed.
This loss requires computing the skeletons of both the ground truth and the predicted segmentation in a differentiable manner.
Currently, differentiable skeletonization algorithms are either inaccurate or computationally demanding.
In this paper, we propose to use a U-Net which computes the vascular skeleton directly from the segmentation and the MRA image. This method is naturally differentiable and provides a good trade-off between accuracy and computation time.
The resulting cascaded multitask U-Net is trained with the clDice loss to embed topological constraints during the segmentation. 
In addition to this topological guidance, this cascaded U-Net also benefits from the inductive bias generated by the skeletonization during the multitask training.
This model is able to predict both the vessel segmentation and centerlines with a more accurate topology and a low computation time.
Code is available at : \url{https://github.com/PierreRouge/Cascaded-U-Net-for-vessel-segmentation}
\keywords{vessel segmentation, U-Net, topological loss, deep-learning.}
\end{abstract}
\section{Introduction}
\label{sec:intro}

Vascular diseases include all the alterations that occur in vessels, such as stenosis, aneurysm, thrombosis or embolism.
Consequences of these diseases, such as stroke, are an important cause of death and disability worldwide.
The accurate vessel segmentation from angiographic images, actively investigated over the last 30 years \cite{DBLP:journals/mia/LesageABF09,DBLP:journals/cmpb/MocciaMHM18}, is an important step for the diagnosis and treatment of vascular diseases. 

In the last decade, deep-learning has allowed significant progress in medical imaging \cite{Shen:ARBE:2017} and especially in segmentation.
In particular, it is increasingly applied to the segmentation of vascular structures \cite{mookiah2021review}. 
In this context, one of the first approaches dealing with 3D angiographic images was proposed by Kitrungrotsakul et al. \cite{kitrungrotsakul_vesselnet_2019} who used a multi pathway convolutional network processing each of the three orthogonal planes to segment the hepatic vessels in 3D.
The same year, Livne et al. \cite{livne_u-net_2019} and Sanches et al. \cite{DBLP:conf/isbi/SanchesMVN19} used for the first time architectures building upon 2D and 3D U-Net, respectively, to segment the brain vessels. 
From this time on, efforts were geared towards designing networks dedicated to the segmentation of curvilinear structures. 
Notably, Mou et al. \cite{mou2021cs2} proposed a convolutional network with two attention modules to encode both spatial and channel relationship. 
This method was tested on six different imaging modalities and nine datasets (2D and 3D).
In the meantime, Tetteh et al. \cite{tetteh2020deepvesselnet} designed an architecture aiming to perform simultaneously vessel segmentation, centerline prediction and bifurcation detection in angiographic images.
Recently, Dang et al. \cite{dang2022vessel} tackled the problem of data annotation and proposed a weakly-supervised deep-learning framework.
The annotated patches were obtained using a classifier discriminating between vessel and non-vessel patches and K-means algorithm.

Despite these efforts, automatic segmentation of vascular networks remains a challenging topic, especially due to the complex topological and geometrical properties of vessels, and their sparseness in the images.
By contrast to many anatomical structures, vessels do not constitute a compact volume at a specific position and scale. 
They are organized as a multiscale network (from large vessels to thin ones, close to the resolution of the image) in the whole image.
This represents a challenge for deep-learning methods, especially when a topologically correct result is required for subsequent tasks such as blood flow modeling \cite{Miraucourt:CMBB:2017}.

To overcome these difficulties, Shit et al. \cite{shit2021cldice} recently proposed a novel metric specifically designed to evaluate the quality of tubular structure segmentation.
This metric, named clDice (for ``centerline Dice''), mainly relies on the skeleton of the tubular structures instead of their whole volume, therefore focusing on topological information. 
To use this new metric as a loss function, it is necessary to compute the skeleton of the predicted segmentation in a differentiable manner.
Therefore, the authors proposed a differentiable soft-skeleton algorithm. However, the resulting skeletons do not preserve the topology of the structures of interest.
In a following work, Menten et al. \cite{menten2023skeletonization} proposed two new differentiable skeletonization algorithms to overcome this limitation.
They showed that using the clDice loss with this soft-skeleton algorithm provides better and more connected segmentation of 2D tubular structures, for instance on retinal images and on 3D tubular structures on the Vessap dataset \cite{todorov2020machine}, a dataset of mice brain vascular networks acquired at a very high resolution and with a research protocol.
However, such approach has not yet been tested on Magnetic Resonance (MR) or X-ray Computed Tomography (CT) angiographic dataset acquired in clinical conditions (\textit{i.e.} images with more noise, artifacts and with a lower resolution).

\begin{figure*}[t]
    \centering
    \includegraphics[width=1\linewidth]{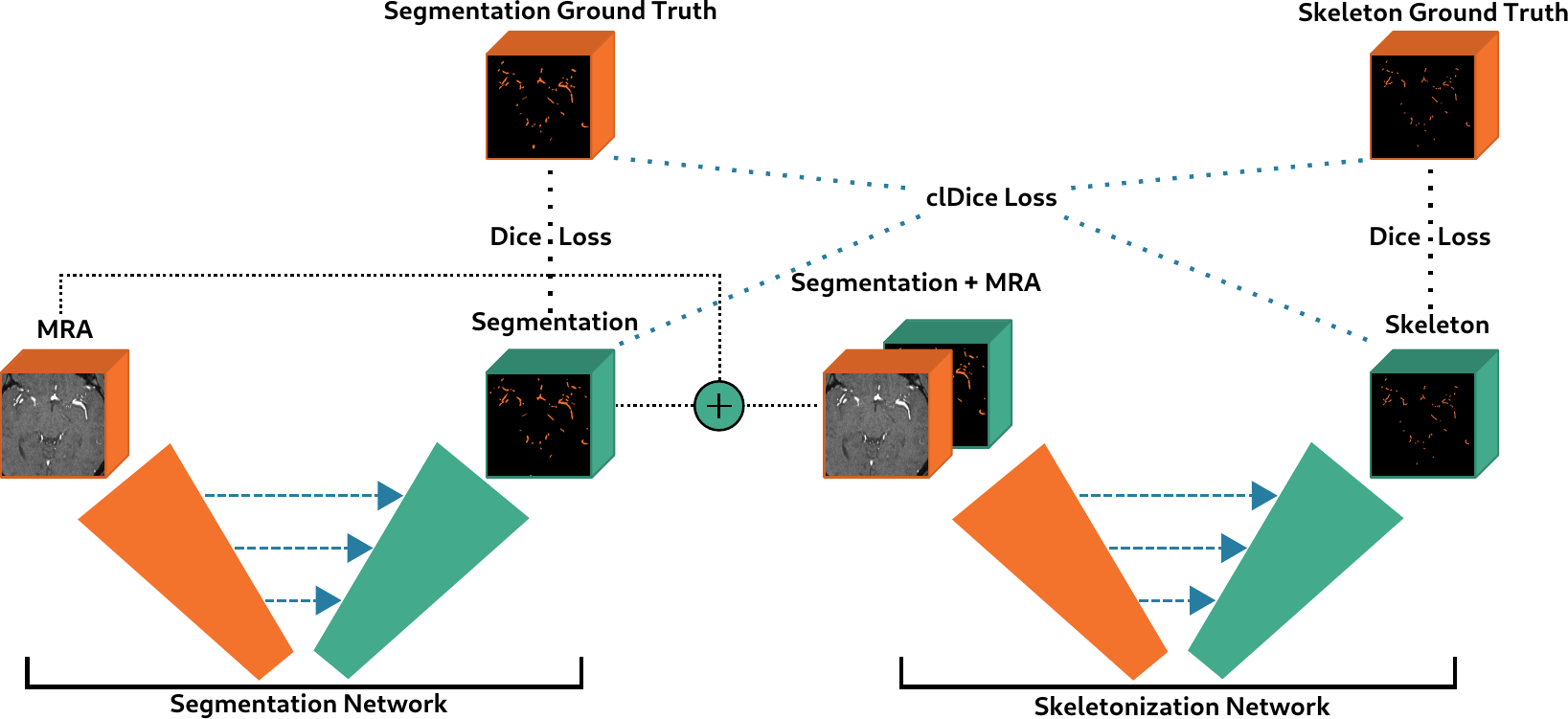}
    \caption{Architecture of the proposed cascaded U-Net (see Section~\ref{sec:model}).}
    \label{fig:architecture_wnet}
\end{figure*}

Our work is inspired by Pan et al. \cite{pan2021msc} who proposed to perform the skeletonization via a multitask architecture with shared features fusion blocks which jointly perform centerline extraction and vessel segmentation.
Here, the skeleton is computed based on the initial image and benefits only indirectly from the segmentation. 

In this work, we propose a cascaded network with a U-Net backbone \cite{ronneberger2015u} that first computes the segmentation, and then uses this segmentation and the initial image to perform the skeletonization task. In this architecture, the skeletonization tasks directly benefits from the segmentation, and may also incorporate information from the initial image to produce a better skeleton.
Finally, the skeleton output, obtained in a differentiable manner, will be used to compute the clDice loss and supervise the whole network, resulting in a more topologically correct segmentation.

We evaluate our method against four standard U-Net, trained with the Dice or clDice losses using the skeletonization methods introduced in \cite{shit2021cldice} and \cite{menten2023skeletonization}.
We show that our method provides segmentations with a more accurate topology and with a lower computation time.

In this context, our main contributions are the following: 
\begin{itemize}
    \item we evaluate the performance of the clDice loss for 3D vascular segmentation of the human brain;
    \item we propose an efficient way of performing the skeletonization operation to compute the clDice;
    \item we propose a cascaded multitask U-Net architecture which segments brain vascular network with a more accurate topology and with a lower runtime.
\end{itemize}
The remainder of the article is organized as follows.
In Section~\ref{SEC:Method}, we describe our methodological contribution.
In Section \ref{SEC:Evaluation}, we present experiments that assess the relevance of our method.
In Section \ref{SEC:Conclusion}, we discuss our results and conclude.

\section{Method}
\label{SEC:Method}

\begin{table*}[!t]
    \centering
    \caption{Results of our skeletonization network vs. others skeletonization methods}
    \begin{tabular}{l c c c c}
         \hline
         Model & Runtime (ms)  $\downarrow$ & $\chi$ error $\downarrow$ & $\beta_0$ error $\downarrow$ & $\beta_1$ error $\downarrow$\\
         \hline
         soft-skeleton algorithm & $5$ $\pm$ $1$ &  $1313$ $\pm$ $257$ & $1168$ $\pm$ $241$ & $144$ $\pm$ $30$ \\
         Euler & $558$ $\pm$ $13$ & $1$ $\pm$ $1$ & $2$ $\pm$ $2$ & $1$ $\pm$ $1$\\ 
         Boolean & $1022$ $\pm$ $34$ & $1$ $\pm$ $1$ & $1$ $\pm$ $1$ & $0$ $\pm$ $0$\\
         \hline
         Skeletonization network & $9$ $\pm$ $2$ & $412$ $\pm$ $69$ & $294$ $\pm$ $48$ & $118$ $\pm$ $26$\\
         \hline
    \end{tabular}
    \label{tab:res_skeleton}
\end{table*}

\begin{table*}[!t]
    \centering
        \caption{Results of the Cascaded Multitask U-Net: mean $\pm$ standard deviation values ($p$-value).
        The $p$-values are computed with respect to the U-Net (Dice) results.}
    \begin{tabular}{lrrrrrr}
             \hline
         Model & $DSC \uparrow$  & $clDice \uparrow$ &   $\chi$ error $\downarrow$ & $\beta_0$ error $\downarrow$ & $\beta_1$ error $\downarrow$ & Runtime (s) $\downarrow$\\
         \hline
         U-Net (Dice) & $\textbf{0.76}$ $\pm$ $0.02$ & $\textbf{0.85}$ $\pm$ $0.02$ & $37.8$ $\pm$ $20.3$ & $9.5$ $\pm$ $6.5$ & $33.3$ $\pm$ $24.4$ \\
         U-Net (Dice + clDice Soft) &  $0.75$ $\pm$ $0.02$ & $\textbf{0.85}$ $\pm$ $0.02$ & $33.7$ $\pm$ $19.3$ & $8.2$ $\pm$ $6.8$ & $30.0$ $\pm$ $22.6$ & $128$ $\pm$ $1$\\
           U-Net (Dice + clDice Euler) &  $0.75$ $\pm$ $0.02$ & $\textbf{0.85}$ $\pm$ $0.02$ & $26.6$ $\pm$ $15.2$ & $7.8$ $\pm$ $6.0$ & $\textbf{25.0}$ $\pm$ $17.4$ &  $198$ $\pm$ $1$\\
            U-Net (Dice + clDice Boolean) & $\textbf{0.76}$ $\pm$ $0.01$ & $\textbf{0.85}$ $\pm$ $0.02$ & $24.5$ $ \pm$ $15.9$ & $\textbf{7.0}$ $\pm$ $6.7$ & $25.8$ $\pm$ $16.8$ &  $301$ $\pm$ $1$\\
         \hline
       Cascaded U-Net (ours) & $0.75$ $\pm$ $0.02$ & $0.84$ $\pm$ $0.02$ & $ \textbf{23.5}$ $\pm$ $14.8$ & $\textbf{7.0}$ $\pm$ $6.7$ & $ 25.5$ $\pm$ $16.5$ & $170$ $\pm$ $1$\\        
      $p$-value w.r.t U-Net (Dice) & $p < 0.05$ & $p=0.3$ &$p < 0.05$  & $p < 0.05$ & $p < 0.05$ \\
      \hline
    \end{tabular}
    \label{tab:res_cascaded_unet}
\end{table*}

\subsection{clDice loss and differentiable skeletonizations algorithms}
\label{SSEC:clDice loss and soft-skeleton algorithm}

The clDice \cite{shit2021cldice} derives from two metrics called \textit{topology precision} ($T_{prec}$) and \textit{topology sensitivity} ($T_{sens}$) in reference to the usual precision and sensitivity metrics.
These metrics are defined as follows:
\begin{equation*}
    T_{prec}(C_P,S_{G}) = \frac{|C_P \cap S_{G}|}{|C_P|} \quad T_{sens}(C_{G},S_P) = \frac{|C_{G} \cap S_P|}{|C_{G}|}
\end{equation*}
where $C_P$, $C_{G}$ and $S_P$, $S_{G}$ are the predicted and ground truth centerlines and segmentations, respectively.
The clDice is defined as the harmonic mean of $T_{prec}$ and $T_{sens}$:
\begin{equation}
    clDice(S_P,S_{G}, C_P, C_{G}) = 2 \cdot \frac{T_{prec}(C_P,S_{G}) \cdot T_{sens}(C_{G},S_P)}{T_{prec}(C_P,S_{G})+T_{sens}(C_{G},S_P)} \nonumber
\end{equation}

By leveraging the skeleton representation, the clDice avoids being biased by large vessels and thus better focuses on topological information. 
Most of the methods designed to extract a skeleton are not differentiable. 
Therefore, Shit et al. proposed a differentiable soft-skeleton algorithm to use the clDice for training a neural network.
This algorithm uses min and max filters to perform dilation and erosion on the predicted segmentation.
Preliminary experiments (see Section \ref{ssec:Skeletonization using U-Net}) showed that the results from this soft-skeletonization are not sufficiently accurate for 3D vascular segmentation, in particular regarding topology.

In a following work, Menten et al. \cite{menten2023skeletonization} proposed two new differentiable skeletonization algorithms.
These algorithms remove simple points \cite{DBLP:journals/tmi/SahaSB15} in the image, ensuring that the topology is not affected by the skeletonization.
The identification of simple points is done either using the Euler characteristics or through a Boolean characterization. In the following, we will then refer to these methods as Euler and Boolean methods, respectively.
These methods have the advantage of generating a nearly perfect skeleton in terms of topology, but at the cost of an important computation time.

In our work, we chose to use a standard U-Net to perform the skeletonization. This method is, by nature, differentiable and presents a good trade-off between accuracy and computation time.

\subsection{Model architecture}
\label{sec:model}

The backbone model used is a standard U-Net \cite{ronneberger2015u} with a depth of $4$, using $2$-stride convolution for down-sampling, instance normalization and leakyReLU activation function.



Our cascaded U-Net architecture is presented in Figure~\ref{fig:architecture_wnet}.
It is composed of a first U-Net taking as input an MRA image and performing the segmentation. This task is supervised by a Dice loss and will be referred to as the \textit{segmentation network}.
The output of this network is concatenated with the MRA image and fed to a second U-Net performing the skeletonization task, also supervised by a Dice loss.
This part of the architecture will be referred to as the \textit{skeletonization network}.
Similarly to Pan et al. \cite{pan2021msc}, the training of these two networks is also supervised by the clDice loss.
The cascaded U-Net final loss is then defined by:
\begin{multline}
    Loss(S_P,S_{G}, C_P, C_{G}) = Dice(S_P,S_{G}) + \\
    \lambda_1 \cdot Dice(C_P,C_{G}) +\lambda_2 \cdot clDice(S_P,S_{G}, C_P, C_{G})
\end{multline}
where $\lambda_1, \lambda_2 \in \mathbb R$ are two weight parameters.

This architecture presents several advantages. First, the skeletonization network performs the skeletonization in a differentiable manner, which allows using the clDice loss and enforcing topological constraint on the segmentation task.
In addition, unlike the other skeletonization methods, the skeletonization network, that takes as input both the output of the segmentation network and the original MRA image, can correct segmentation errors and therefore limits error propagation.
Finally, by jointly learning segmentation and skeletonization, the segmentation network will benefit of an inductive bias that will encourage learning topologically correct segmentations.
Indeed, the skeletonization task gives similar importance to all vessels independently of their thickness; so the multitask learning can help to enforce the importance of small vessels in the segmentation task.  

\subsection{Training configuration}
\label{SSEC:Training configuration}

All the MRA volumes were first normalized by Z-score.
During training, one batch is composed of $2$ patches of size $192\times192\times64$, each randomly located in an MRA volume. One epoch consists of $250$ batches.

To ensure a fair comparison across our experiments, we used the same data augmentation strategy for all trained networks, which is directly inspired by nnUnet \cite{isensee2021nnu}.



Sliding windows with $25\%$ overlap were used to reconstruct the full volume at inference time.
A Gaussian kernel was applied on overlapping parts to reduce the weight of the voxels far from the center of the patches.
We used stochastic gradient descent with Nesterov momentum with an initial learning rate set to $0.01$.
A linear learning rate decay was applied so that learning rate was equal to $0$ at the last epoch.

For the cascaded multitask U-Net, we chose to first pretrain the two U-Nets separately before fine-tuning the network in the scheme described in Section~\ref{sec:model}.
We initialized the weights of the segmentation network with the weights of the U-Net trained with the Dice loss, and we pretrained the skeletonization network using as input the MRA image and the segmentation ground truth.
Baseline models were trained during 750 epochs; for the cascaded multitask U-Net we first pretrained the two networks during $750$ epochs, then fine-tuned it during $250$ epochs.


\section{Evaluation}
\label{SEC:Evaluation}

In this section, we present the setup and results of the experiments conducted to evaluate our skeletonization network and cascaded multitask U-Net.
All the results presented for deep-learning models were obtained through a $5$-fold cross-validation.

\begin{figure*}[!t]
    \centering
    \includegraphics[width=1\linewidth]{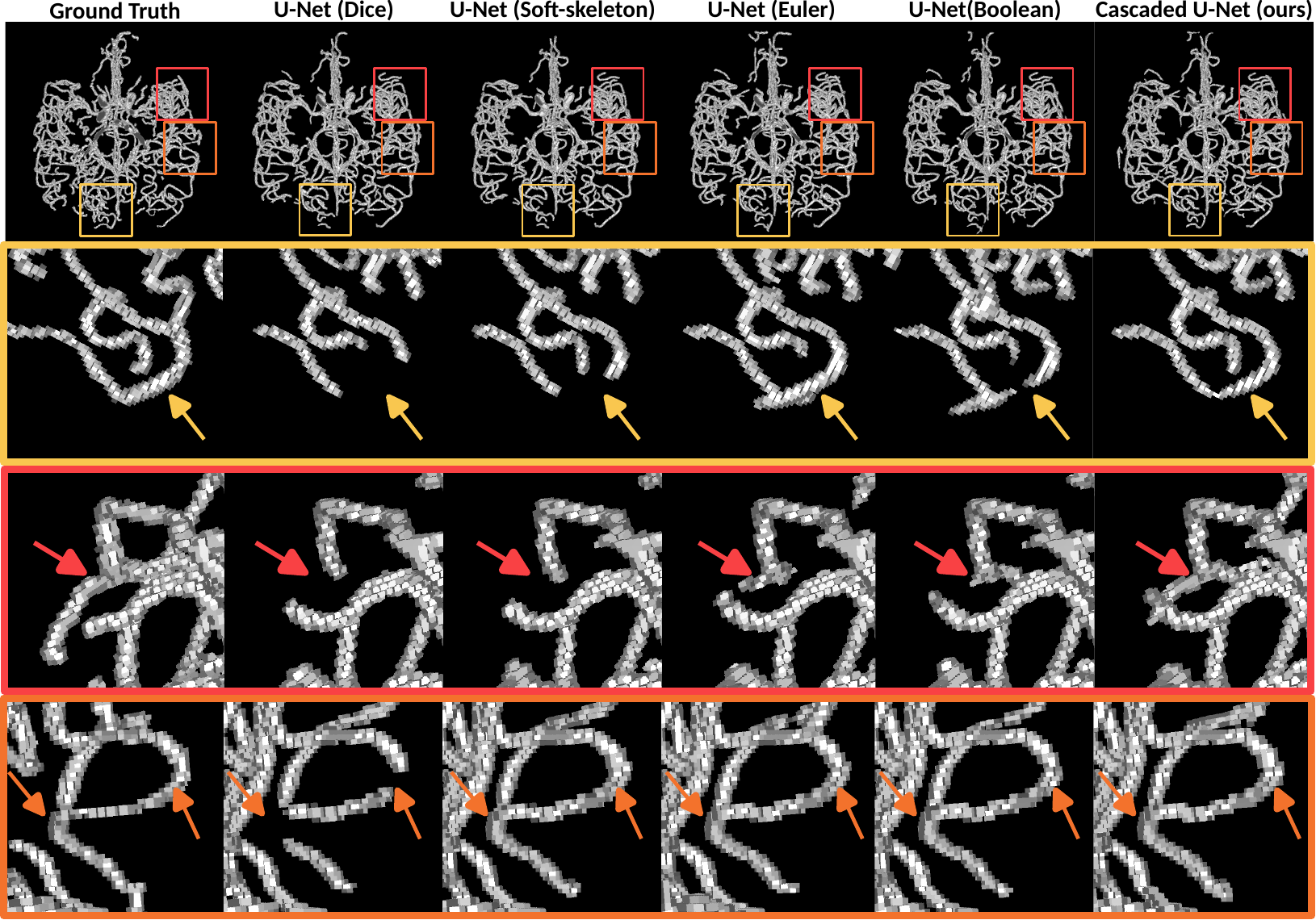}
    \caption{Segmentation results obtained with the different approaches for patient 8. Color boxes indicate the corresponding zoomed area.}
    \label{fig:visual_result}
\end{figure*}

\subsection{Dataset and metrics}
\label{sec:dataset}

\textbf{Dataset:} For this study, we used a publicly available dataset\footnote{\url{https://public.kitware.com/Wiki/TubeTK/Data}} \cite{aylward2002initialization} containing $34$ time-of-flight (TOF) MRA volumes of the brain.
All the volumes present a voxel resolution of $0.513 \times 0.513 \times 0.800$ mm$^{3}$ and a size of $448 \times 448 \times 128$.
For all the volumes, the associated vessel segmentation ground truth was available.
To produce the skeleton ground truth, we used the \texttt{skeletonize} function from the python package \texttt{scikit-image}.\\

\noindent \textbf{Metrics:} In the evaluation, we used clDice (see Section~\ref{SSEC:clDice loss and soft-skeleton algorithm}) and Dice similarity coefficient ($DSC$).
Beyond these quantitative metrics, we also evaluated the topological quality of our segmentation by computing the following topological scores: mean absolute error of the first two Betti numbers $\beta_0$ (number of connected components) and $\beta_1$ (number of tunnels), and mean absolute error of the Euler characteristics $\chi = \sum_i (-1)^i\beta_i$ with $\beta_i$ the successive Betti numbers.
Before computing the metrics, we applied a post-processing on all the results to remove the connected components with a volume lower than $100$ voxels. 

\subsection{Skeletonization using U-Net}
\label{ssec:Skeletonization using U-Net}

To evaluate our skeletonization network, we compared it to the soft-skeleton algorithm introduced in \cite{shit2021cldice}, the skeletonization based on the Boolean characterization of simple points and the skeletonization based on the Euler Characteristic introduced in \cite{menten2023skeletonization}.
We computed, for all methods, the mean time required to perform the skeletonization per patch as well as the topological metrics introduced in Section~\ref{sec:dataset}.
These results, presented in Table~\ref{tab:res_skeleton}, show that our method produces better skeletons than the soft-skeleton algorithm with regard to the topological metrics in a comparable computation time. 
Euler and Boolean methods produce nearly perfect skeletons but at the cost of a more important runtime.
Our simple skeletonization network then provides a good trade-off between accuracy and computation time.
Qualitatively, the skeletons produced by the soft-skeleton algorithm present many disconnections and a thickness of several voxels compared to other skeletonization methods.
By contrast, our method provides more accurate skeletons.


\subsection{Cascaded Multitask U-Net}

\subsubsection{Hyperparameters optimization}
The goal of the cascaded multitask U-Net is to improve the results of the segmentation network thanks to the clDice loss.
As stated in Section~\ref{sec:model}, we have to tune the hyperparameters $\lambda_1$ and $\lambda_2$ in order to handle the trade-off between the skeletonization loss and the clDice loss.
We also tested two training configurations: a first one where the skeletonization network weights are frozen; a second where they are updated during the cascaded U-Net training.
In both configurations, the skeletonization network was first pre-trained.
Based on these experiments, we found that the best cascaded multitask U-Net training policy consists of freezing the weights of the skeletonization network and set the loss weights to $(\lambda_1, \lambda_2) = (0.5, 0.5)$.

\subsubsection{Validation and comparisons}

We compared the segmentation of our cascaded U-Net with four methods: 
\begin{itemize}
    \item a U-Net trained with a Dice loss;
    \item a U-Net trained with the combination of Dice loss and clDice loss but with the soft-skeleton, Euler or Boolean algorithms.
\end{itemize}
For all these methods, the trade-off between Dice and clDice loss was optimized through a grid search.

Table~\ref{tab:res_cascaded_unet} presents the results of the different methods.
All methods present similar results in terms of Dice and clDice.
However, we can observe that every method using the clDice loss succeeds to improve the topological correctness of the segmentations, as indicated by the improvement of the topological metrics. This behavior is also illustrated in Figure~\ref{fig:visual_result}.

Besides, we can see that the model using soft-skeleton algorithm presents the worst topological metrics, confirming that the quality of the skeletonization is an important factor to use the clDice loss. 
Our method succeeds to give a similar topological improvement with respect to the models using the Euler and Boolean skeletonization methods but with an improved runtime (average epoch duration).
We attribute the good behavior of our method---despite an imperfect skeletonization---to two factors. First, unlike deterministic methods, a learned skeletonization network can correct segmentation error during training (notably using the information of the MRA image) and therefore limits error propagation. Second, in our method, the skeletonization acts as a complementary task. Therefore, thanks to this joint multitask learning, the segmentation task can benefit of an inductive bias that helps to learn a topologically correct segmentation of its own.


\section{Conclusion}
\label{SEC:Conclusion}

In this article, we proposed to use a U-Net to learn the skeletonization operation required to compute the clDice loss. This method provides a good trade-off between the topological correctness of the skeletons and the computation time. 
We then proposed a cascaded multitask U-Net to learn a vascular segmentation with a topological constraint enforced by the clDice loss.
This cascaded multitask U-Net jointly learns vessel segmentation and skeletonization and can then benefit of the inductive bias induced by the skeletonization task.

In this study, we showed that the clDice loss improves the topological correctness of vascular segmentation from MRA images.
Moreover, we demonstrated that our method gives the best topological correctness with the least computation time.
Our method may certainly be improved by investigating dedicated skeletonization networks or improved information sharing between the segmentation and skeletonization tasks.
Such improvements will be part of future works.

\section{Compliance with ethical standards}

This research study was conducted retrospectively using human subject data made available in open access by Kitware at the following link: \linebreak \url{https://public.kitware.com/Wiki/TubeTK/Data}. Ethical approval was not required as confirmed by the license attached with the open access data

\section{Acknowledgements and conflicts of interests}

This work was supported by the French Agence Nationale de la Recherche (Grant ANR-20-CE45-0011).
This work was granted access to the HPC resources of IDRIS under the allocation 2022-AD011013610 made by GENCI.

The authors have no relevant financial or non-financial interests to disclose.

%
%
%
\bibliographystyle{splncs04}
%
\bibliography{biblio}

\end{document}